\documentclass[galaxies,article,accept,pdftex,10pt,a4paper]{mdpi} 
\firstpage{1} 
\articlenumber{x}
\doinum{10.3390/------}
\pubvolume{xx}
\pubyear{2016}
\copyrightyear{2016}
\externaleditor{Academic Editor: }
\history{Received: ; Accepted: ; Published: }
\pdfoutput=1

 \theoremstyle{mdpi}
 \newcounter{thm}
 \newcounter{ex}
 \newcounter{re}
 \theoremstyle{mdpidefinition}

\Title{Time-dependent injection as a model for rapid blazar flares }

\Author{Michael Zacharias $^{1}$}
\AuthorNames{Michael Zacharias}

\address[1]{%
$^{1}$ Landessternwarte, Universit\"at Heidelberg, D-69117 Heidelberg, Germany; m.zacharias@lsw.uni-heidelberg.de}

\corres{Correspondence: m.zacharias@lsw.uni-heidelberg.de; Tel.: +x-xxx-xxx-xxxx}

\abstract{Time-dependent injection can cause non-linear cooling effects, which lead to a faster energy loss of the electrons in jets. The most obvious result is the appearance of unique breaks in the SED, which would normally be attributed to a complicated electron distribution. The knowledge of the observation time and duration is important to interpret the observed spectra, because of the non-trivial evolution of the SED. Intrinsic gamma-gamma absorption processes in the emission region are only of minor importance.}


\conferencetitle{Blazars through sharp multiwavelength eyes}

\begin{document}



\section{Introduction}
Blazars are active galaxies, where the line-of-sight is closely aligned with the relativistic jet. Due to the strong Doppler boost of the produced radiation, the physics of the jet can be probed with great precision. The standard model invokes emission by highly relativistic electrons (in the frame of the jet) interacting with the magnetic field and ambient photon fields, such as the synchrotron photons (the synchrotron-self Compton (SSC) process), or external photon fields originating from the accretion disk, the broad-line region, the dusty torus, or the CMB. A review is given by, e.g., \cite{b07}.

The observation of minute-short flares in blazars \cite{aea07,Aea07} challenges the standard one-zone model, which is usually invoked to explain blazar emission. The variability time scale implies either an extremely small emission region, which is in contradiction with the high luminosities recorded from these flares, or an extremely high Doppler factor, which is in contradiction to radio observations of moving knots in the jets of blazars \cite{hs06}.

Several models have been developed to overcome these problems. These include the jet-in-a-jet model \cite{gea09}, the similar mini-jets-in-a-jet model \cite{bg12}, jet-star interactions \cite{bea12}, and others. These have in common that they invoke an emission region, which is smaller, denser, and faster than the surrounding jet material. Such features could develop from magnetic reconnection events within the jet.

{Reconnection events might result in a time-dependent injection of particles into the radiation zone.} This changes the cooling of the electrons beyond the standard behavior. In fact, sources can enter a parameter space where the electron cooling becomes collective causing nonlinearities and time-dependencies. { Here, an analytical description of this effect is given, which} has profound implications for the resulting spectral energy distributions (SEDs) and lightcurves. Since internal absorption of photons is similar to the standard model, time-dependent injection should be considered as an important addition to the models explaining rapid variability in blazars.

\section{Time-dependent SSC cooling}

The kinetic equation describing the electron distribution function $n(\gamma,t)$ due to cooling and injection is given by \cite{k62}
\begin{eqnarray}
 \frac{\partial n(\gamma,t)}{\partial t} - \frac{\partial}{\partial \gamma}\left[ |\dot{\gamma}_{tot}| n(\gamma,t) \right] = q_0 \delta(\gamma-\gamma_0) \delta(t), \label{eq:kineq}
\end{eqnarray}
where $\gamma$ is the electron Lorentz factor. The injection is modeled as a single burst of particles with energy $\gamma = \gamma_0$ at time $t=0$. Since no further injection takes place, the electron distribution cannot reach an equilibrium state, and the particles will continue to cool. The cooling term $\dot{\gamma}_{tot}$ contains contributions from synchrotron cooling, cooling on the external photons, and cooling from the SSC process. Thus,
\begin{eqnarray}
 |\dot{\gamma}_{tot}| =& |\dot{\gamma}_{syn}| + |\dot{\gamma}_{ec}| + |\dot{\gamma}_{ssc}| \nonumber \\
 =& D_0 (1+l_{ec}) \gamma^2 + A_0 \gamma^2 \int\limits_0^{\infty} \gamma^{\prime 2} n(\gamma^{\prime},t) \mbox{d}\gamma^{\prime} \label{eq:elcool},
\end{eqnarray}
where $D_0$ and $A_0$ are constants depending on the source parameters, such as the magnetic field and the radius. The synchrotron and external Compton cooling term have been combined, since both { keep the differential equation linear}. The relative strength between these two processes is given by the parameter
\begin{eqnarray}
 l_{ec} = \frac{|\dot{\gamma}_{ec}|}{|\dot{\gamma}_{syn}|} =  \frac{4\Gamma_b^2}{3} \frac{u_{ec}^{\prime}}{u_B} \label{eq:lec},
\end{eqnarray}
with the bulk Lorentz factor $\Gamma_b$, the magnetic energy density $u_B$, and the energy density in external photons measured in the galactic frame $u_{ec}^{\prime}$.

The SSC cooling term is nonlinear and time-dependent due to the integral over the electron distribution \cite{s09}. Since we only take into account the cooling of the particles, the strength of the SSC cooling decreases with time, since the energy density stored in the electrons decreases with time. Therefore, even for strong initial SSC cooling, after some time the SSC cooling will become weaker than the linear coolings causing a change in the cooling behavior. This can be quantified by the injection parameter, given as
\begin{eqnarray}
 \alpha = \sqrt{\frac{|\dot{\gamma}_{ssc}(t=0)|}{|\dot{\gamma}_{syn}|+|\dot{\gamma}_{ec}|}} = \sqrt{\frac{A_0q_0}{D_0(1+l_{ec})}}\gamma_0 \label{eq:alpha}
\end{eqnarray}
which characterizes the initial conditions of the source. Obviously, for increasing $q_0$ and $\gamma_0$ the source is more SSC dominated, while for stronger external fields the source is more linearly dominated. It should be noted that $\alpha$ is independent of the magnetic field strength.

The solution for Eq. (\ref{eq:kineq}) has been calculated by \cite{sbm10} and depends on the value of $\alpha$. For $\alpha\ll1$, i.e. linearly cooling sources, the solution becomes
\begin{eqnarray}
 n(\gamma,t) = q_0 \delta\left( \gamma - \frac{\gamma_0}{1+D_0(1+l_{ec})\gamma_0 t} \right) \mbox{H}\left[ \gamma_0-\gamma \right] \label{eq:sol01}.
\end{eqnarray}
For initial nonlinear cooling, $\alpha\gg 1$, the solution is divided into two parts, namely
\begin{eqnarray}
 n(\gamma,t<t_c) =& q_0 \delta\left( \gamma - \frac{\gamma_0}{(1+3\alpha^2 D_0(1+l_{ec})\gamma_0 t)^{1/3}} \right) \mbox{H}\left[ \gamma_0-\gamma \right] \label{eq:sol10a}, \\
 n(\gamma,t\geq t_c) =& q_0 \delta\left( \gamma - \frac{\gamma_0}{\frac{1+2\alpha^3}{3\alpha^2}+D_0(1+l_{ec})\gamma_0 t} \right) \mbox{H}\left[ \frac{\gamma_0}{\alpha}-\gamma \right] \label{eq:sol10b}.
\end{eqnarray}
At the critical time
\begin{eqnarray}
 t_c = \frac{\alpha^3-1}{3\alpha^2 D_0(1+l_{ec})\gamma_0} \label{eq:tc}.
\end{eqnarray}
the cooling behavior changes, with strong implications on the emerging spectra and lightcurves.

The injection of a power-law in energy gives similar results, since the power-law is rapidly quenched into a near-delta-like distribution \cite{zs10}. 

\section{The spectral energy distribution}

\begin{figure}
\begin{minipage}[t]{0.48\textwidth}
\centering \resizebox{\hsize}{!}
{\includegraphics{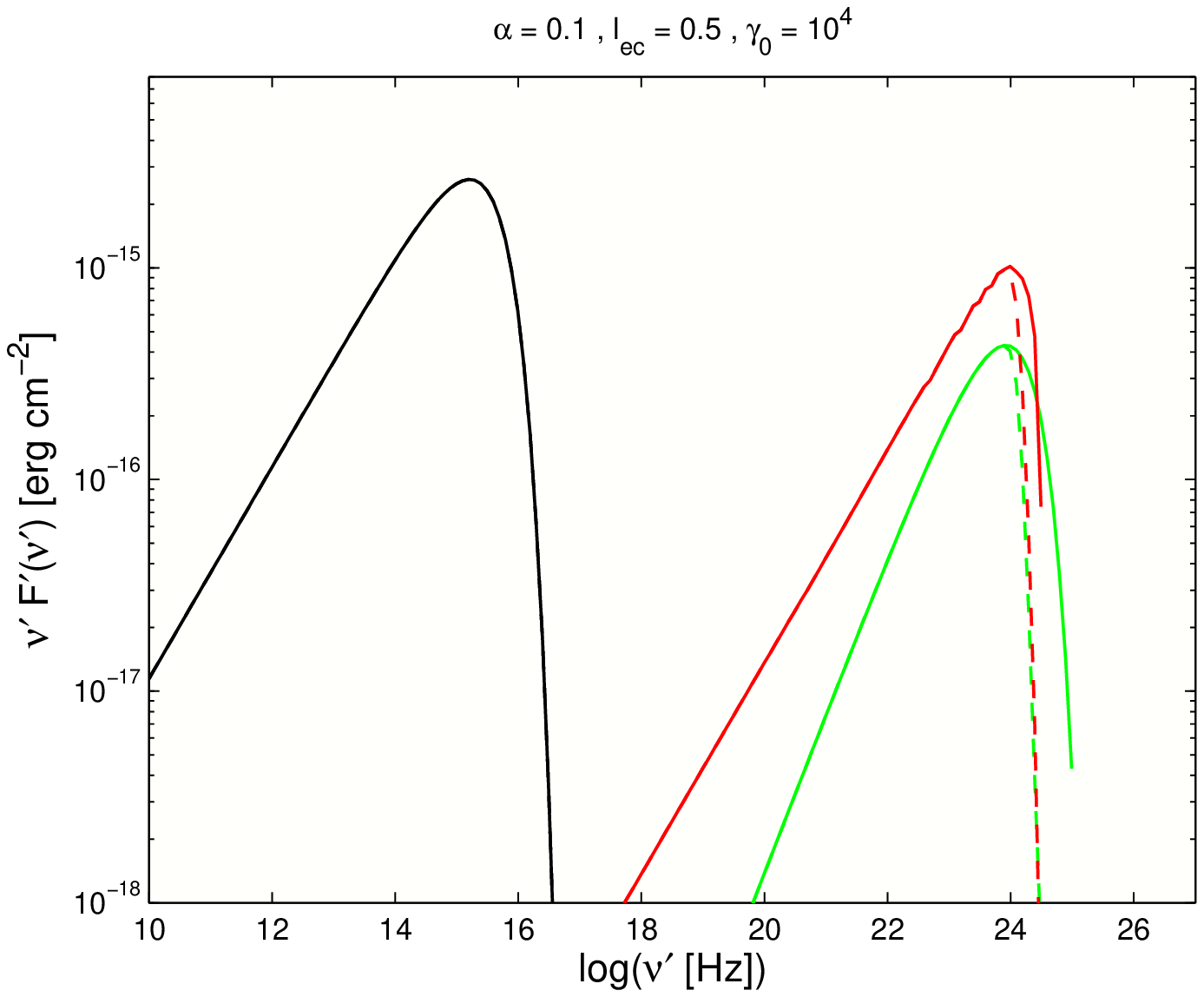}}
\end{minipage}
\hspace{\fill}
\begin{minipage}[t]{0.48\textwidth}
\centering \resizebox{\hsize}{!}
{\includegraphics{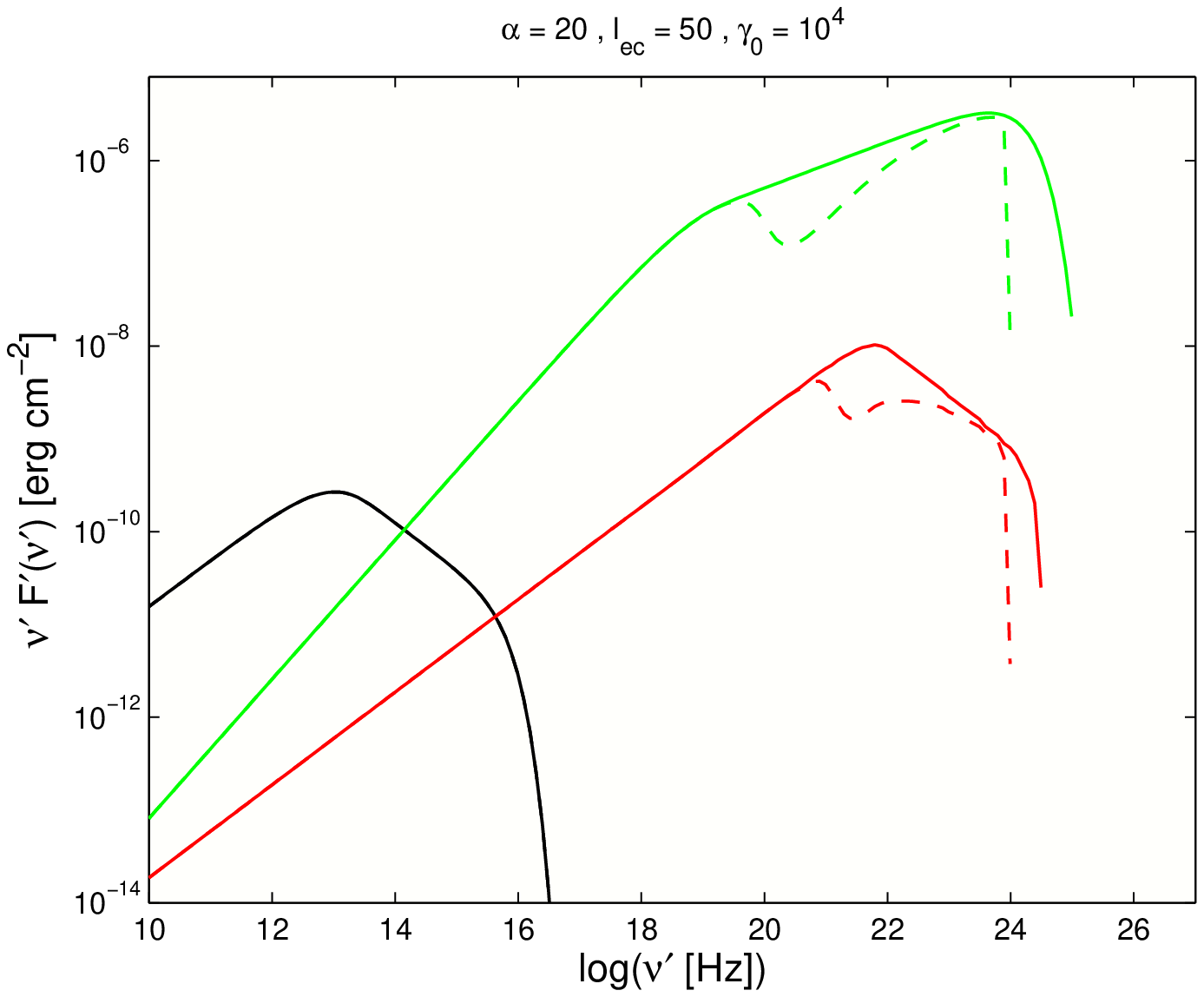}}
\end{minipage} 
\caption{Total SEDs with synchrotron (black), SSC (green), and external Compton (red). Dashed lines with internal and external absorption. \textit{Left:} SED for $\alpha\ll 1$ and $l_{ec}\ll 1$. \textit{Right:} SED for $\alpha\gg 1$ and $l_{ec}\gg 1$. Note the different scales on the y-axis.}
\label{fig:sed}
\end{figure} 
The intensity of a given process (namely, synchrotron, SSC, and external Compton) is calculated by folding the electron distribution function with the respective emission power. The SED is then derived by a time-integration of the intensity. Details for all components can be found in \cite{sbm10,zs12a,zs12b}.

Integrating from $0\rightarrow\infty$ in time gives the total SED presented as the solid lines in Fig. \ref{fig:sed}. The left plot shows the result for $\alpha\ll 1$, while the right plot shows the result for $\alpha\gg 1$. As expected, the cooling strongly determines the dominating component. For $\alpha\ll 1$ the SEDs of all components exhibit a single power-law followed by an exponential cut-off, which is the typical result for an injection of delta-function in energy, even if the injection is continuous in time. Thus, time-dependent and continuous injections are indistinguishable under linear cooling.

For $\alpha\gg 1$ all components exhibit a broken power-law, where the position of the break strongly depends on $\alpha$. For example, the synchrotron component breaks by unity, which is often observed in blazars, but cannot be explained by a typical ``cooling break'' which usually gives a break of $0.5$. This also shows that a broken power-law with a stronger break than $0.5$ in the SED can be explained without the need for complicated electron distributions.

Since the intensity is time-dependent, an observed SED strongly depends on the observation time and the duration. It is therefore difficult to model non-simultaneous data, since the different observation times must be taken into account. This can be understood, if one considers the time-dependent evolution of the theoretical SED. Thus, we have created two short movies, which show the evolution and the gradual build-up of the SED over time. These can be accessed at the publicly available domain:
\begin{itemize}
 \item $\alpha\ll 1$: \url{www.tp4.rub.de/~mz/SEDa01.mp4}
 \item $\alpha\gg 1$: \url{www.tp4.rub.de/~mz/SEDa10.mp4}
\end{itemize}
The movies illustrate how the total SED is gradually built up over time. It is also evident that the nonlinear cooling sets in quicker than the linear cooling, and how the break in the SED is created once the cooling switches from nonlinear to linear. One can also observe that the SSC SED evolves faster than the synchrotron and external Compton SEDs, since the electron distribution influences the SSC process twice (creation of synchrotron photons and subsequent scattering).

\section{Photon-photon absorption}

The dashed lines in Fig. \ref{fig:sed} indicate the absorption of photons by the pair creation process. The absorption at high $\gamma$-energies is caused by the stationary external photon field.

Internal absorption has the interesting feature of being time-dependent. Since the intensities are time-dependent, the photons necessary for photon-photon pair creation are not available at all times in the required numbers to create a significant effect. This implies that the source is optically thick for different energies at different times. Hence, the source is optically thin for the highest $\gamma$-ray energies at early times and becomes optically thick for them at later times. However, due to the cooling, no more high energy $\gamma$-rays are produced, and the effect is unobservable. Similarly, for the photons produced at the low energy end of the Compton components the source is optically thick at the beginning of the flare. However, at these times only a very small portion of these photons are produced. At later times, when the most photons at these are energies are created, the source is optically thin.

There is only an intermediate regime, where the source can be optically thick at a time, when the majority of these photons are created. These energies reveal themselves by the trough in the SED of Fig. \ref{fig:sed}. This effect is stronger for more extreme parameters. Unfortunately the interesting energy regime is not covered currently by any observatory, so this feature cannot be observed at this time.

Details concerning the derivation and interpretation can be found in \cite{z15}.

\section{The lightcurves}

\begin{figure}
\begin{minipage}[t]{0.48\textwidth}
\centering \resizebox{\hsize}{!}
{\includegraphics{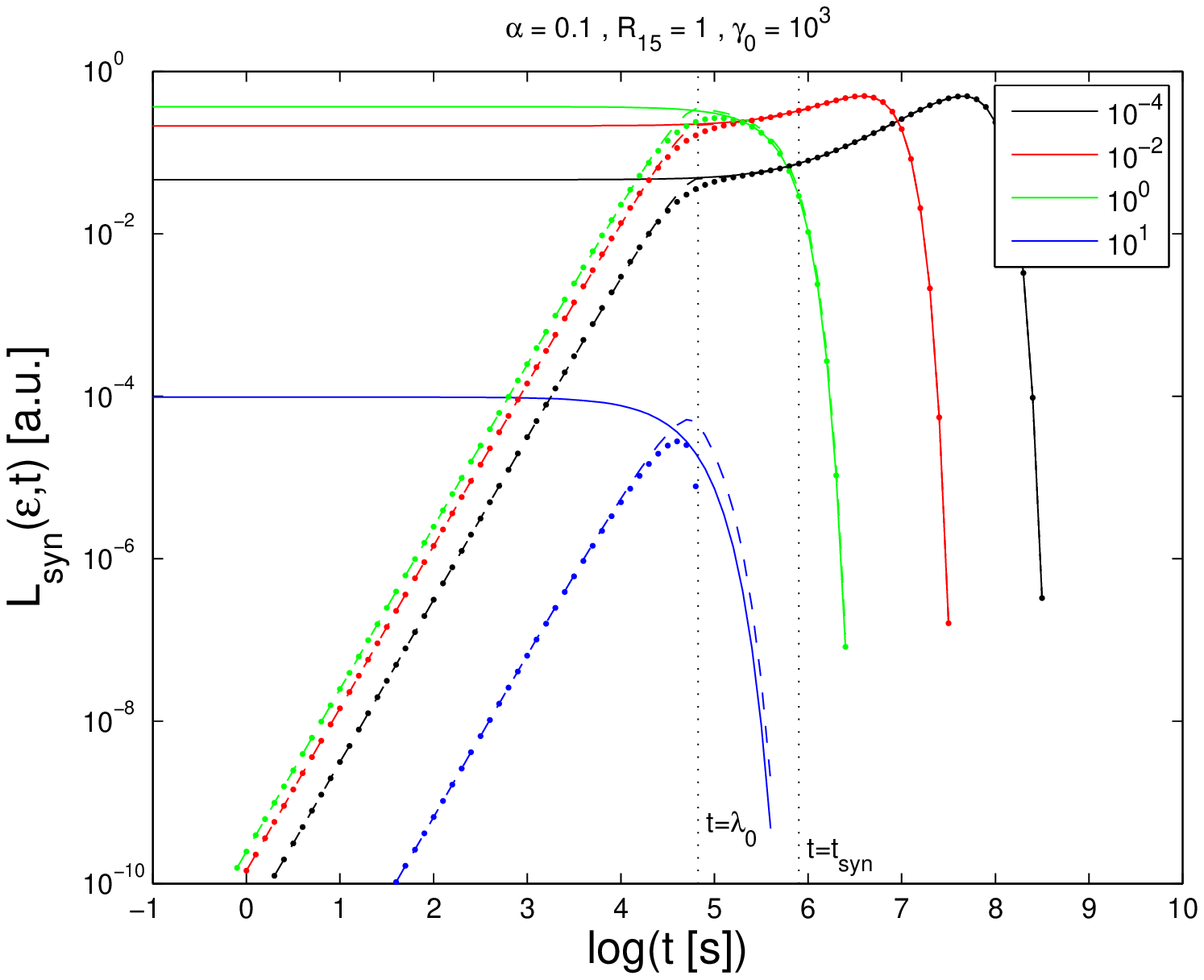}}
\end{minipage}
\hspace{\fill}
\begin{minipage}[t]{0.48\textwidth}
\centering \resizebox{\hsize}{!}
{\includegraphics{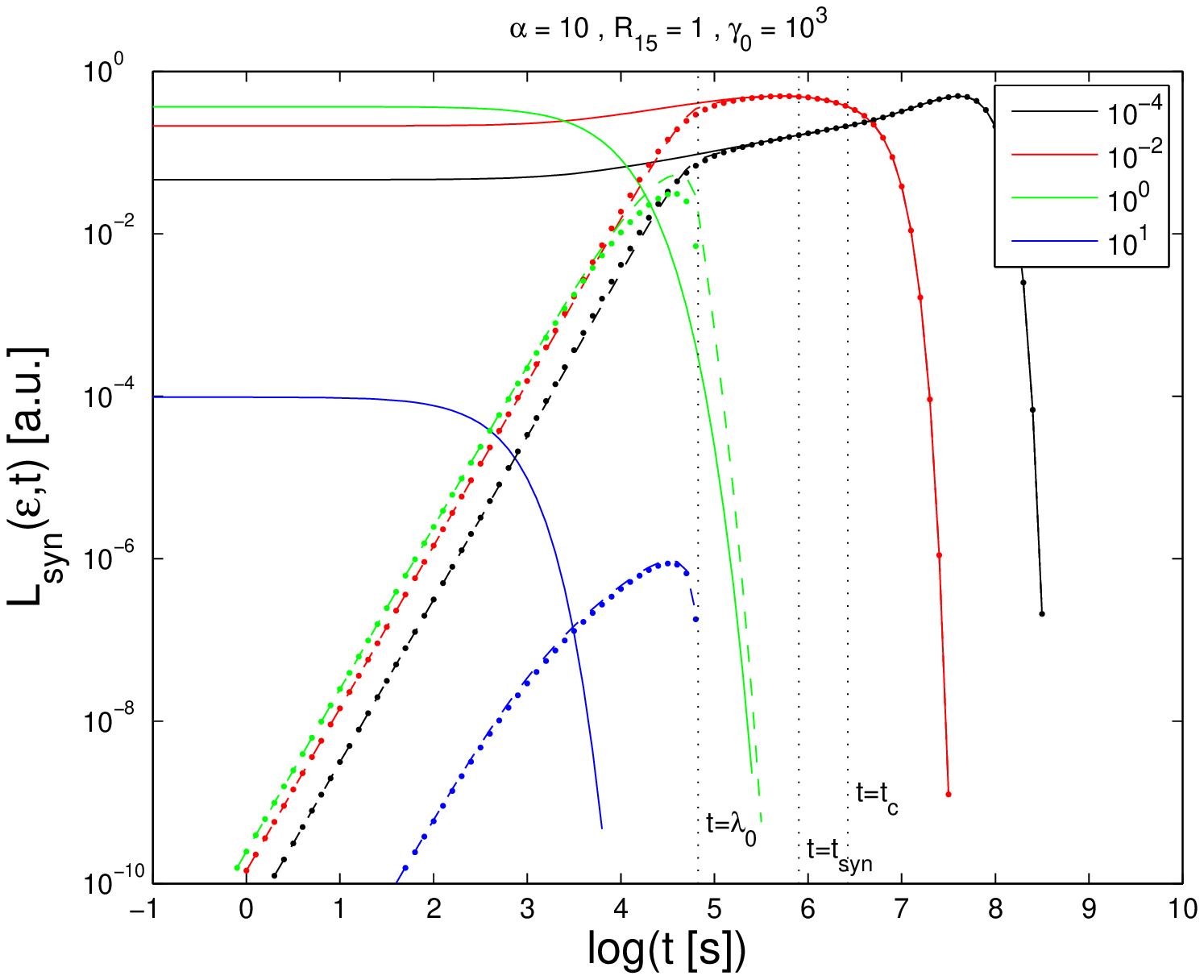}}
\end{minipage} 
\caption{Synchrotron lightcurves for two cases of $\alpha$. Displayed are the lightcurves without retardation (solid), numerically integrated lightcurves with retardation (dashed), and analytically integrated lightcurves with retardation (dotted). The energies are normalized to the synchrotron energy of maximum flux.}
\label{fig:lc}
\end{figure} 
Fig. \ref{fig:lc} shows the exemplary synchrotron lightcurves for the two cases of $\alpha$ for four different energies. The dashed and dotted lines include the effect of photon retardation as described in \cite{zs13}. As can be seen, compared to the lightcurve without the retardation effect the signal builds up during the initial time until the lightcrossing time $\lambda_0 = 2R/c$ has passed. After that time, the retardation effect becomes unimportant.

The retardation washes out the prominent effects of the nonlinear cooling especially at the highest energies (blue and green curve). However, at lower energies the effects are still visible as can be seen by comparing the red and the black curve in the respective cases of $\alpha$.

For the SSC and external Compton lightcurves the results are similar, as is discussed in \cite{z14}.

\section{Summary and Conclusions}

We have discussed the effect of time-dependent injection on the cooling process and the influence on the observed spectra and lightcurves. Due to the time-dependent injection the electron distribution cannot reach the equilibrium state. Thus, the nonlinear nature of the SSC cooling becomes important resulting in a time-dependent cooling behavior. This has two major implications. Firstly, the cooling acts quicker than in purely linear scenarios, so that the ``half-energy'' cooling time is much reduced. Secondly, the cooling behavior changes, because the SSC cooling strength decreases with time and at some point the linear cooling takes over. { The resulting effects have been studied in the limiting case of a delta-function injection in both energy and time, which gives the opportunity to calculate analytically the entire SED and the lightcurves.}

The SED is strongly influences by this change in cooling behavior, because it causes breaks in the SED, which are stronger than the usual cooling break. It is important to note that these breaks do not depend on the injected electron distribution, and therefore do not require any complicated electron distribution often invoked to explain such strong SED breaks.

Furthermore, in a time-dependent scenario, the SED depends strongly on the time of observation. Therefore, only strictly simultaneous data should be compared with a single model curve. Otherwise the properties of the source might have changed dramatically.

The internal photon-photon absorption is not very important, unless the source enters a rather extreme parameter space. Due to the time-dependency of the intensity distribution, the optical thickness of the source is also time-dependent, and mostly affects intermediate Compton energies located roughly around the MeV domain.

The lightcurves are strongly influenced by the photon retardation at early times washing out the effects of the fast nonlinear cooling. Only at low energies, which evolve slower than the high energies, the effect of the nonlinear cooling becomes visible. The retardation effect implies that the nonlinear model faces the same difficulties as the usual one-zone model, namely that the variability timescale is given by the lightcrossing time.

In conclusion, time-dependent injection and, thus, time-dependent cooling gives very interesting results for both the SED and the lightcurves. { Especially, the large-$\alpha$ case, where SSC cooling dominates initially, should be considered for modeling rapid flaring events in blazars. It could explain the change in the Compton dominance observed during flares of some sources \citep[e.g.,][]{aea09}.}

\vspace{6pt} 


\acknowledgments{I wish to thank Reinhard Schlickeiser for a fruitful collaboration, especially during my time as his Ph.D. student. Support by the German Ministry for Education and Research (BMBF) through Verbundforschung Astroteilchenphysik grant 05A11VH2 is gratefully acknowledged. }

\appendixtitles{no} 
\appendixsections{multiple} 
%

\bibliographystyle{mdpi}

\renewcommand\bibname{References}



\end{document}